\documentclass[hyperref,twocolumn,showpacs,preprintnumbers,amsmath,amssymb,superscriptaddress]{revtex4}

\usepackage{graphicx}
\usepackage{dcolumn}
\usepackage{bm}
\usepackage[usenames]{color}
\usepackage{epsfig}
\usepackage{amssymb,array}
\def \irbaddress{Rudjer Bo\v{s}kovi\'{c} Institute, Bijeni\v{c}ka cesta 54, P.O. Box 180, 10002 Zagreb, Croatia}
\def \acuaddress{Abilene Christian University, ACU Station Box 7963, Abilene, TX 79699, USA}
\def \untzaddress{University of Tuzla, Faculty of Science, Univerzitetska 4, 35000 Tuzla, Bosnia and Herzegovina}
\begin{document}

\title{Multichannel Anomaly of the Resonance Pole Parameters Resolved}
\author{Sa\v sa Ceci}
\email{sasa.ceci@irb.hr}
\affiliation{\irbaddress}
\author{Jugoslav Stahov}
\affiliation{\acuaddress}
\affiliation{\untzaddress}
\author{Alfred \v{S}varc}
\affiliation{\irbaddress}
\author{Shon Watson}
\affiliation{\acuaddress}
\author{Branimir Zauner}
\affiliation{\irbaddress}

\date{\today}

\begin{abstract}
Inspired by anomalies which the standard scattering matrix pole-extraction procedures have produced in a mathematically well defined coupled-channel model, we have developed a new method based solely on the assumption of partial-wave analyticity. The new method is simple and applicable not only to theoretical predictions but to the empirical partial-wave data as well. Since the standard pole-extraction procedures turn out to be the lowest-order term of the proposed method the anomalies are understood and resolved.
\end{abstract}

\maketitle

\paragraph*{Introduction.}
The determination of the scattering matrix (S-matrix) is considered to be the major objective of both, scattering theory and energy-dependent analysis of scattering data. The collection of S-matrix poles in the ``unphysical'' Riemann sheet is related to resonance mass spectrum \cite{New82,BraMor} so obtaining them is the crucial goal of any partial-wave analysis. There is, however, a long lasting (and yet unresolved) controversy on the resonances' physical properties. It is not clear whether physical mass and decay width of a resonance are given by the ``conventional'' resonance parameters like Breit-Wigner mass and the decay width, or by resonance pole parameters---real part and \mbox{$-2\times$} imaginary part of pole \cite{Hoe93,Wor99}. In the case of baryon resonances, the compromise is achieved in a way that all the conventional, as well as pole parameters are collected in the Review of Particle Physics (RPP) \cite{PDG}.

The motivation for this work came from the fact that pole parameters we extracted from our coupled-channel partial waves \cite{Bat98} by a standard model-independent pole searching method (speed plot \cite{Hoe93}) were not unique. The obtained pole positions varied from one reaction to another. The time-delay \cite{Kel04}, another method of choice for resonance pole extraction was employed, but with similar outcome. This anomalous behavior challenged common sense and the conclusion was drawn that either our partial-wave analysis or the applied pole extraction methods were incorrect. The extraction methods were carefully examined, and those methods were determined to be at fault. This effort resulted in a new model-independent extraction method free from this anomaly. In addition, we calculated elastic pole residues, and the obtained values were in quite good agreement with others published in the RPP \cite{PDG}. Since there are no RPP estimates for elastic residues yet, this result lends support to provide them.

It is important to draw a distinction between the time-delay signal as resonance's ``smoking gun'', and utilization of the time-delay (or speed plot) for the extraction of resonance parameters. The authors agree that former plays a central role as one of resonance existence criteria. However, we disagree profoundly with the interpretation of the latter as proper resonance parameter extraction method. In fact, we show that the speed plot is just the lowest-order approximation to a correct pole extraction method.

In this Letter we introduce two methods for obtaining the resonance pole parameters from energy-dependent partial waves. One is based on analytic continuation of a channel propagator. The analytic continuation happens to be dependent on the analysis model, since the channel propagator is a feature of the Carnegie-Mellon-Berkeley (CMB) analysis \cite{Cut79}. The other, T-matrix ``regularization'' (i.e. pole eliminating) method is formulated in the model-independent manner, so pole extraction is not restricted to CMB formalism. Moreover, the only required information may be attained from measurable physical processes. Therefore we recommend it as The Method for obtaining resonance pole parameters.

The N* pole parameters given in this Letter were extracted from partial waves obtained in our current partial-wave analysis \cite{Bat98}.

\paragraph*{The Core of CMB Approach.} Our current partial-wave analysis \cite{Bat98} is based on the CMB approach \cite{Cut79}. The most prominent property of this approach is analyticity of partial waves with respect to Mandelstam $s$ variable. In every discussion on partial-wave poles, analyticity plays a crucial role since poles are situated in a complex plane, away from physical region. Any knowledge about the nature of partial-wave singularities would be impossible to gain if partial waves were not analytic functions. The ability to calculate pole positions is not just a benefit of the model's analyticity but also a necessity for the CMB resonance extraction. In this approach, the resonance itself is considered to exist if there is an associated partial-wave pole in the ``unphysical'' sheet.

The central role of the CMB analysis belongs to the unitary-normalized partial-wave T-matrix $\mathbf{T}(z)$ \cite{Cut79,Bat98,Vra00}. It is a matrix in channel indices, and generic complex variable $z$ in this section denotes Mandelstam $s$. The connection between S- and T-matrix is given by \mbox{$\bm{S}(z)=\bm{I}+2\,i\,\bm{T}(z)$} where $\bm{I}$ is the unit matrix. Two main ingredients of the model are channel propagator $\mathbf{\Phi}(z)$ - the diagonal matrix in channel indices which takes care of channel related singularities, and bare resonant propagator $\mathbf{G}_0(z)$ - the diagonal matrix in resonant indices incorporating real first-order poles related to resonances (and background). Background contribution is given by two sub-threshold poles (another pole may be placed further above considered energy region). Dressed channel propagator $\mathbf{G}(z)$ is given by the resolvent (Bethe-Salpeter) equation
\mbox{$\bm{G}^{-1}(z)=\bm{G}_0^{-1}(z)-\bm{\Sigma}(z)$},
where self-energy term $\bm{\Sigma}(z)$ is built from the channel propagator as \mbox{$\bm{\gamma}\cdot\bm{\Phi}(z)\cdot\bm{\gamma}^\mathrm{T}$}. The parameter matrix $\bm{\gamma}$ is a non-square matrix obtained from the least-square fit to experimental or partial-wave data. In addition to $\bm{\gamma}$ parameter matrices, the values of the bare propagator real poles are concurrently acquired from the same fit. The partial-wave data are fitted by the unitary-normalized partial-wave T-matrix given by relation
\begin{equation}\label{eq:Tmatrix}
\mathbf{T}(z)=\sqrt{\mathrm{Im}\,\mathbf{\Phi}(z)}\cdot\bm{\gamma}^\mathrm{T}\cdot\mathbf{G}(z)\cdot\bm{\gamma}\cdot\sqrt{\mathrm{Im}\,\mathbf{\Phi}(z)} .
\end{equation}

Channel propagator matrix $\mathbf{\Phi}(z)$ is assembled from channel propagator functions $\phi(z)$. The dominant singularity in the resonant region, apart from resonances themselves, is the physical (channel opening) branching point $x_o$. In the CMB approach, contributions from other singularities (left-hand cut, nucleon pole etc.) are given partly by the design of the channel propagator imaginary part, while the rest is taken care of by the background.

Analyticity of channel propagator function $\phi(z)$ is ensured by once-subtracted dispersion relation
\begin{equation}\label{eq:DR}
\phi(z)=\frac{z-x_o}{\pi}\,\,\mathrm{P}\int_{x_o}^{\infty}\frac{\mathrm{Im}\,\phi(x')\,dx'}{(x'-z)(x'-x_o)},
\end{equation}
where $\mathrm{P}$ stands for Cauchy principal value. The physical (unitarity) branch cut is, thus, chosen to go from the branching point $x_o$ to positive infinity. The variable $x'$ is used in the integral rather than $z'$ to indicate the integration path is on the real axis.

The form of the channel propagator imaginary part is given as
\begin{equation}\label{eq:imphi}
\mathrm{Im}\, \phi(x)=\frac{\left[q(x)\right]^{2L+1}}{\sqrt{x}\,\left\{Q_1+\sqrt{Q_2^2+\left[q(x)\right]^2}\right\}^{2L}},
\end{equation}
where $q(x)$ is the standard two-body center of mass momentum for a particular meson-baryon channel, $Q_1$ and $Q_2$ are model parameters with values equal to the $\pi$ meson (or, in our case, the channel meson \cite{Bat98}) mass. $L$ is the orbital angular momentum number of the given partial wave.

\paragraph*{Extraction Method One: Analytic Continuation.} From Eq.\ (\ref{eq:imphi}) it is evident that $\phi(z)$ has a square-root type singularity. Instead of calculating the dispersion integral (\ref{eq:DR}) for each point in complex plane, we decided to use the expansion (similar to Pietarinen's in Ref. \cite{Pie75})
\begin{equation}\label{eq:expansion}
\phi_{\mathrm{I}}(z)=\sum_{n=0}^{N}c_n\,\left(Z_{\mathrm{I}}(z)\right)^n,
\end{equation}
where $c_n$ are coefficients of expansion. The new channel dependent variable is given by its principal branch
\begin{equation}
Z_{\mathrm{I}}(z)=\frac{\alpha-\sqrt{x_o-z}}{\alpha+\sqrt{x_o-z}},
\end{equation}
with the tuning parameter $\alpha$. This function is fitted to a dataset consisting of imaginary parts of $\phi(x)$ from Eq.\ (\ref{eq:imphi}) and real parts of $\phi(x)$ calculated from dispersion relation (\ref{eq:DR}), both of them evaluated at real axis (hence $x$). The general idea is that the $\phi(z)$ inherits analytic structure from $Z(z)$. We obtained parameters $\alpha$ and coefficients $c_n$ for each channel, and for all analyzed partial waves. The least-square fit is considered to be good if it meets following conditions: (i) small number of coefficients $c_n$ needed (7 or 8, at most), (ii) the function fitted to the part of dataset, when extrapolated outside of the fitted region is consistent with the rest of data, and (iii) fitting just imaginary part of $\phi(x)$ produces real part that is in agreement with values obtained from (\ref{eq:DR}).

The channel propagator given by expansion (\ref{eq:expansion}) is obtained quite accurately and works very well in the resonant region in the vicinity of physical axis. 

Every channel opening is responsible for two Riemann sheets: the first (physical) sheet with physical partial waves, and the secondary (unphysical) sheet with resonant poles. To get to the unphysical sheet it is enough to use the second branch of $Z(z)$
\begin{equation}\label{eq:simplerecipe}
Z_{\mathrm{II}}(z)=\frac{\alpha+\sqrt{x_o-z}}{\alpha-\sqrt{x_o-z}}.
\end{equation}

Finally, it is evident from Eq. (\ref{eq:Tmatrix}) that all poles of each partial wave must be by construction the same in every channel and, in fact, equal to poles of the resolvent $\mathbf{G}(z)$. Here we use T-matrices obtained in our last partial-wave analysis \cite{Bat98} and collect all poles of $\mathbf{G}(z)$ obtained by analytic continuation in \mbox{Table \ref{tbl:poles}}. Since this method gives poles of partial waves in Mandelstam $s$ variable, comparison to RPP estimates is made with the square root of the Mandelstam pole (selecting branch with positive real part) denoted by $\mu$.

\begin{table*}
\caption{The $N^*$ resonance pole parameters obtained by two methods from this Letter along with RPP \cite{PDG} estimates. The N/E is written if a resonance pole position does not have RPP estimate, while the $N(????)$ stands for resonances unnamed in the RPP. For the $N(1535)$ S$_{11}$ resonance, the extracted parameters were strongly influenced by the singularity of the $\eta N$ channel opening so the minimizing process was onerous.}
\label{tbl:poles}
\begin{ruledtabular}
\begin{tabular}{lcllcccccc}
\multicolumn{4}{c}{\textsc{Review of Particle Physics} \cite{PDG}} & \multicolumn{2}{c}{\textsc{Analytic Cont.}}  & \multicolumn{4}{c}{\textsc{Regularization Method}} \\
\multicolumn{1}{c}{N*} & L$_{2I\,2J}$& \multicolumn{1}{c}{$\mathrm{Re}\,\mu$} & \multicolumn{1}{c}{$-2\,\mathrm{Im}\,\mu$} &   \multicolumn{1}{c}{$\mathrm{Re}\,\mu$} & \multicolumn{1}{c}{$-2\, \mathrm{Im}\,\mu$} &  \multicolumn{1}{c}{$\mathrm{Re}\,\mu$} & \multicolumn{1}{c}{$-2\, \mathrm{Im}\,\mu$} & \multicolumn{1}{c}{$|r|$} & \multicolumn{1}{c}{$\,\theta$}\\
& & \multicolumn{1}{c}{(MeV)} & \multicolumn{1}{c}{(MeV)} & \multicolumn{1}{c}{(MeV)} & \multicolumn{1}{c}{(MeV)} & \multicolumn{1}{c}{(MeV)} & \multicolumn{1}{c}{(MeV)} &\multicolumn{1}{c}{(MeV)} & \multicolumn{1}{c}{($^\circ$)}\\ 
\cline{1-4}\cline{5-6}\cline{7-10}\\
$N(1535)$ &S$_{11}$ & 1505(10) & 170(80)  & 1517 & 190   & 1522 & 146 & 19 & $-$146 \\
$N(1650)$ &S$_{11}$ & 1660(20) & 160(10)  & 1642 & 203  & 1647 & 203 & 84 & $-$58\\
$N(2090)$ &S$_{11}$ & \multicolumn{1}{c}{N/E}& \multicolumn{1}{c}{N/E}& 1785 & 420 & - & - & - & - \\ 
$N(1440)$ &P$_{11}$ & 1365(20) & 210(50)  & 1359 & 162    & 1354 & 162 & 47 & $-$95 \\
$N(1710)$ &P$_{11}$ & 1720(50) & 230(150) & 1728 & 138    & 1729 & 150 & 52 & $-$156\\
$N(????)$ &P$_{11}$ & \multicolumn{1}{c}{N/E}&\multicolumn{1}{c}{N/E}& 1708 & 174  & - & - & - & -\\
$N(2100)$ &P$_{11}$ &\multicolumn{1}{c}{N/E}&\multicolumn{1}{c}{N/E}& 2113 & 345  & 2120 & 347 & 31 & $-$59\\ 
$N(1720)$ &P$_{13}$ & 1700(50) & 250(140) & 1686 & 235    & 1691 & 235 & 19 & $-$112\\ 
$N(1520)$ &D$_{13}$ & 1510(5) & 115(5)    & 1505 & 123    & 1506 & 124 & 36 & $-$14 \\
$N(1700)$ &D$_{13}$ & 1680(50) & 100(50)  & 1805 & 130    & 1806 & 132 & 7 & $-$36\\
$N(2080)$ &D$_{13}$ & \multicolumn{1}{c}{N/E}&\multicolumn{1}{c}{N/E}& 1942 & 476  & - & - & - & - \\ 
$N(1675)$ &D$_{15}$ & 1660(5) & 140(15)   & 1657 & 134    & 1658 & 138 & 25 & $-$20\\
$N(2200)$ &D$_{15}$ & \multicolumn{1}{c}{N/E}&\multicolumn{1}{c}{N/E}& 2133 & 437  & 2145 & 439 & 22 & $-$71\\ 
$N(1680)$ &F$_{15}$ & 1670(5) & 120(15)   & 1664 & 134   & 1666 & 136 & 45 & $-$26\\ 
$N(1990)$ &F$_{17}$ & \multicolumn{1}{c}{N/E}&\multicolumn{1}{c}{N/E}& 1990 & 303  & 2016 & 318 & 8 & $-$25\\ 
$N(????)$ &G$_{17}$ &\multicolumn{1}{c}{N/E}&\multicolumn{1}{c}{N/E}& 1740 & 270   & 1749 & 280 & 6 & $-$86\\
$N(2190)$ &G$_{17}$ & 2050(100) & 450(100) & 2060 & 393    &   2068 & 389  & 34 & $-$30\\
\end{tabular}
\end{ruledtabular}
\end{table*}

\paragraph*{The Anomaly Appears.} It is well known that each resonance pole on the unphysical sheet is accompanied with poles on other Riemann sheets \cite{Ede64,BraMor,Mor87}. They are attached to the same resonance phenomena, but with dissimilar pole positions as well as elastic pole residues \cite{Cut90}. Pole parameters presented in the RPP \cite{PDG} are taken from the unphysical Riemann sheet (the one closest to the physical sheet). To be sure that the simple recipe given by Eq.\ (\ref{eq:simplerecipe}) provides us with proper pole parameters (i.e. that we are searching for poles on the right sheet), we compared results with those obtained by standard model-independent pole extraction methods. The two  renowned extraction methods are speed-plot \cite{Her60,Hoe93,Han96} and time-delay \cite{BraMor,Kel04}. Both methods rely on following parameterization of T-matrix elements (Breit-Wigner)
\begin{equation}\label{eq:BWparameterization}
T(z)=\underbrace{\,\,\,\frac{r}{\mu-z}\,\,\,}_{\mathrm{resonant\, part}}+\,\,\underbrace{\left(T(z)-\frac{r}{\mu-z}\right)}_{\mathrm{smooth\, background}},
\end{equation}
where $\mu$ and $r$ are pole position and pole residue, respectively. Here $z$ stands for center-of-mass energy ($\sqrt{s}$). If one plots modulus of the ``speed'' of T (i.e. $\left|dT(z)/dz\right|$), the resonance produces a peak in this speed plot. There are, however, known exceptions like $N(1535)$ which is hidden ``under the cloak'' of the $\eta N$ channel opening \cite{Hoe93}. Resonance poles are extracted from Eq.\ (\ref{eq:BWparameterization}) under the erroneous assumption that ``speed'' of the background can be completely neglected when compared to ``speed'' of the resonant part. We compared the speed-plot pole positions of $\pi N$ elastic T-matrix element to positions obtained by using other T-matrix elements: from other quasi-elastic (like $\eta N\rightarrow\eta N$) as well as inelastic processes (e.g. $\pi N\rightarrow\eta N$). Obtained pole positions were shifted by a few tens of MeVs which indicated that there was something wrong.

The time-delay \cite{BraMor,Kel04} is given as the imaginary part of the product of S-matrix inverse and S-matrix ``speed'' 
(\mbox{$\mathrm{Im}\left[\,\bm{S}^\dag (z)\cdot d\bm{S} (z)/dz\right]$}). The resonance reveals itself as peak in a plot of time-delay versus $z$. The parameters are again obtained by neglecting background contributions thus discrepancies that occurred were similar (though less) to those of speed plot.

The background contribution can be disregarded near a pole. However, the real axis may be too far from the pole, so observed discrepancies can be explained by background contribution. Instead of carefully designing this background, we developed a following method to approach the complex pole (where background is negligible) without leaving the real axis.

\paragraph*{Extraction Method Two: Regularization.} Let there be an analytic function $T(z)$ of complex variable $z$ that has a first-order pole at the some complex point $\mu$. This function $T(z)$ can be any T-matrix element, and variable $z$ can be either Mandelstam $s$ or center-of-mass energy $\sqrt{s}$. The latter is used in this method in order to achieve full correspondence with the speed plot. Since all physical processes occur at real energy values, we are allowed to determine directly only values of $T(x)$, where $x$ is a real number. To be able to successfully continue our $T(x)$ into the complex plane to search for its pole(s), we should ``regularize'' this function (i.e. somehow remove the singularity). Then, any simple expansion would converge in the proximity of the removed pole. The appropriate way to ``regularize'' the function with a simple pole at $\mu$, is to multiply it by a form that has a simple zero at the same point \begin{equation} f(z)=(\mu-z)\,T(z). \end{equation} From this definition and Eq. (\ref{eq:BWparameterization}), it is evident that the value of $f(\mu)$ gives the residue $r$ of $T(z)$ at point $\mu$. As we have access to the function values on real axis only, the Taylor expansion of $f$ is done over some real $x$ to give the value (residue) in the pole $\mu$ (where background is highly suppressed)
\begin{equation}
f(\mu)=\sum_{n=0}^N \frac{f^{(n)}(x)}{n!}(\mu-x)^n+R_{N}(x,\mu).
\end{equation}
The expansion is explicitly written to the order $N$, and the rest is designated by $R_{N}(x, \mu)$. The $n$th derivative of $f(x)$ in the form of $T(x)$ is given as
\begin{equation}
f^{(n)}(x)=(\mu-x)\,T^{(n)}(x)-n\,\,T^{(n-1)}(x).
\end{equation}
Insertion of this derivative into Taylor expansion conveniently cancels all consecutive terms in the sum, except the last one
\begin{equation}\label{eq:central}
f(\mu)=\frac{T^{(N)}(x)}{N!}(\mu-x)^{(N+1)}+R_{N}(x,\mu),
\end{equation}
where $T^{(N)}(x)$ is the $N${th} energy derivative of T-matrix element. To simplify the notation, the pole can be written as some general complex number $\mu=a+i\, b$. Once Taylor series converges (i.e. the rest $R_{N}(x,\mu)$ is disregarded) the absolute value of both sides of Eq. (\ref{eq:central}) yields to
\begin{equation}\label{eq:modulus}
\left|f(\mu)\right|=\frac{\left|T^{(N)}(x)\right|}{N!}\left|a+i\,b-x\right|^{(N+1)}.
\end{equation}
To keep the form as simple as possible Eq.\ (\ref{eq:modulus}) is raised to the power of $2/(N+1)$. The elemental second-order polynomial emerges after simple rearrangement on one side of the equation
\begin{equation}\label{eq:finalfit}
\frac{(a-x)^2+b^2}{\sqrt[N+1]{\left|f(\mu)\right|^2}}  = \sqrt[N+1]{\frac{\left(N!\right)^2}{\left|T^{(N)}(x)\right|^2}},
\end{equation}
where the part that is attainable from energy-dependent analysis is put on the right-hand side. The fitting function with just three fit parameters $a$, $b$ and $|f(\mu)|$ is written on the left-hand side. The speed plot method turns out to be identical to the first order of this relation (with \mbox{$N=1$}).

The dataset was produced by the right-hand side of Eq. (\ref{eq:finalfit}). $T^{(N)}(x)$ was given by numerical derivation of energy-dependent partial waves obtained in our analysis \cite{Bat98}. A step of 2 MeV provided a stable procedure. The Taylor expansion is considered to converge when the extracted parameters settle down. The higher orders were used to obtain more accurate values of pole parameters. To acquire reliable fit results, we considered data grouped in a parabolic shape (in accordance with the second-order polynomial).

The elastic-pole residue is commonly given \cite{PDG} by its absolute value $|r|$ and its phase $\theta$, namely
\begin{equation}
|r|=\left|f(\mu)\right| , \,\,\tan\theta={\mathrm{Im}\, f(\mu)}\,/\,{\mathrm{Re}\,f(\mu)},
\end{equation}
where this particular (elastic) $f(\mu)$ is given by the first term in Eq.\ (\ref{eq:central}) with $T(z)$ being $\pi N$ elastic T-matrix element.

Pole parameters attained in this way from $\pi N$ elastic process are given in \mbox{Table \ref{tbl:poles}}. In order to verify the procedure, we applied it to other channel processes. Contrary to anomalous results obtained when using standard procedures, inelastic poles varied by only a few MeV from the elastic ones.

\paragraph*{ Conclusions.} Our new analytic continuation method (Pietarinen expansion of the channel propagator) provides pole positions quickly and precisely while avoiding problems with numerical principal value integration and interpolation. The obtained pole positions are in accordance with RPP values.

The detected anomaly of standard speed plot and time delay methods for pole extraction is fundamental because the unknown contribution (energy derivative of background on the real axis) is considered to be insignificant. The new ``regularization'' method successfully finds resonance pole parameters from a T-matrix in a model independent way. The advantage of the given method is that it can generally be applied to most analytic functions that have a simple pole and values known on any line segment reasonably close to the pole.

The elastic pole residues are in accordance with those given in RPP. Since there are still no residue estimates in RPP, we strongly advocate making them for the future editions.
%



\begin{thebibliography}{Vra00}
\bibitem{New82} R. G. Newton, {\it Scattering Theory of Waves and Particles}, (Springer-Verlag New York, Inc., 1982).
\bibitem{BraMor} B. H. Bransden and R. G. Moorhouse, {\it The Pion Nucleon Systems}, (Princeton University Press, 1973).
\bibitem{Hoe93} G. H\"{o}hler, {\it $\pi N$ Newsletter} {\bf 9}, 1 (1993).
\bibitem{Wor99} R. Workman, Phys. Rev. {\bf C 59}, 3441 (1999).
\bibitem{PDG} S. Eidelman, {\it et al.}, Phys. Lett. {\bf B 592}, 1 (2004).
\bibitem{Bat98} M. Batini\'{c}, {\it et al.}, Phys. Rev. {\bf C 51}, 2310 (1995);
M. Batini\'{c}, {\it et al.}, Physica Scripta {\bf 58}, 15, (1998).
\bibitem{Kel04} N. G. Kelkar {\it et al.}, Nucl. Phys. {\bf A 730}, 121 (2004).
\bibitem{Cut79} R. E. Cutkosky {\it et al.}, Phys. Rev. {\bf D 20}, 2839 (1979).
\bibitem{Vra00} T. P. Vrana, S. A. Dytman and T.-S. H. Lee, Phys. Rep. {\bf 328}, 181 (2000).
\bibitem{Pie75} E. Pietarinen, Il Nuovo Cimento {\bf 12 A} (1972).



\bibitem{Her60} D. J. Herndon, A. Barbaro-Galtieri, and A. H. Rosenfeld, Lawrence Radiation Laboratory Report UCR-20030 $\pi N$, {\it $\pi N$ Partial-Wave Amplitudes - A Compilation}, 1960, p.4, Eq. (9).
\bibitem{Han96} O. Hanstein, D. Drechsel and L. Tiator, Phys. Lett. {\bf B 385}, 45 (1996).


\bibitem{Ede64} R. J. Eden and J. R. Taylor, Phys. Rev. {\bf 133}, B1575, (1964).
\bibitem{Mor87} D. Morgan and M. R. Pennington, Phys. Rev. Lett. {\bf 59}, 2818 (1987).
\bibitem{Cut90} R. E. Cutkosky and S. Wang, Phys. Rev. {\bf D 42}, 235 (1990).
\end{thebibliography}
\end{document}